\documentclass[aps,prb,twocolumn,superscriptaddress,showpacs]{revtex4-1}

\usepackage{graphicx}
\usepackage{amsmath}
\usepackage{amssymb}
\usepackage{indentfirst}
\usepackage{color}

\begin{document}
\title{Structural and electronic properties of Pb$_{1-x}$Cd$_{x}$Te and Pb$_{1-x}$Mn$_{x}$Te ternary alloys}
\author{M. Buka\l{}a}\affiliation{Institute of Physics, Polish Academy of Sciences, Al. Lotnik\'ow 32/46, 02-668 Warsaw, Poland}
\author{P. Sankowski}\affiliation{Institute of Informatics, University of Warsaw, Ul. Banacha 2, 02-097 Warsaw, Poland}
\author{R. Buczko}\affiliation{Institute of Physics, Polish Academy of Sciences, Al. Lotnik\'ow 32/46, 02-668 Warsaw, Poland}
\author{P. Kacman}\affiliation{Institute of Physics, Polish Academy of Sciences, Al. Lotnik\'ow 32/46, 02-668 Warsaw, Poland}

%\email{bukala@ifpan.edu.pl}

\begin{abstract}
A systematic theoretical study of two PbTe-based ternary alloys, Pb$_{1-x}$Cd$_{x}$Te and Pb$_{1-x}$Mn$_{x}$Te, is reported. First, using \emph{ab initio} methods we study the stability of the crystal structure of CdTe - PbTe solid solutions, to predict the composition for which rock-salt structure of PbTe changes into zinc-blende structure of CdTe. The dependence of the lattice parameter on Cd (Mn) content \emph{x} in the mixed crystals is studied by the same methods. The obtained decrease of the lattice constant with \emph{x} agrees with what is observed in both alloys. The band structures of PbTe-based ternary compounds are calculated within a tight-binding approach. To describe correctly the constituent materials new tight-binding parameterizations for PbTe and MnTe bulk crystals as well as a tight-binding description of rock-salt CdTe are proposed. For both studied ternary alloys, the calculated band gap in the L point increases with \emph{x}, in qualitative agreement with photoluminescence measurements in the infrared. The results show also that in p-type Pb$_{1-x}$Cd$_{x}$Te and Pb$_{1-x}$Mn$_{x}$Te mixed crystals an enhancement of thermoelectrical power can be expected.
\end{abstract}

%\keywords{}
\maketitle

\section{Introduction}

Since 1959, when the first Radioisotope Thermoelectric Generator (RTG) was presented, PbTe attracts constantly a lot of interest due to its thermoelectric properties. \cite{Pei,Ref.10,Ref.11} PbTe is also widely used for mid-infrared lasers and detectors. In PbTe the narrow direct band gap at the L point increases from $0.19$ eV at 4.2 K to the value of $E_g = 0.31$ eV at room temperatures.\cite{Dalven} This allows excellent band structure engineering. Recently, also  ternary systems based on lead telluride are of considerable scientific interest because of their potential in device applications.\cite{Ref.1} An important characteristic of the PbTe-based ternary alloys is that their band gap is very sensitive not only to temperature, like in PbTe, but also to composition.  It has been shown that the energy gap of PbTe increases monotonically when alloyed with Cd\cite{Ref.2,Ref.3} as well as with Mn.\cite{Ref.4}

Due to big, ca 1~eV, difference between the energy gaps of PbTe and CdTe, PbTe-CdTe system appears especially suitable for band-gap engineering with a potential for a variety of photonic, thermoelectric and photovoltaic applications. \cite{Ref.5}  The growth of uniform Pb$_{1-x}$Cd$_{x}$Te single crystals is, however, very limited by extremely low mutual solubility of both materials.\cite{Ref.6} The latter results from the difference in the crystal structures -- lead telluride crystallizes in rock-salt structure while cadmium telluride in zinc-blende structure. The limited mutual solubility of PbTe and CdTe was yet exploited for obtaining PbTe quantum dots in a CdTe matrix. This was achieved by thermal annealing of two-dimensional PbTe epilayers embedded in CdTe.\cite{Groiss} In Ref.~\onlinecite{Groiss} it was shown that the size of the dots can be controlled, what  allows for tuning of the quantum dot luminescence over a wide spectral range. As a result, ultrabroadband emission from a multilayered quantum dot stack was demonstrated, which is a precondition for the development of superluminescent diodes operating in the near infrared and midinfrared.

On the other hand, it is expected that thermoelectric properties of PbTe should be improved by implementing CdTe nanostructures in the material. In Ref.~\onlinecite{Hicks} a possibility to increase the thermoelectric figure of merit parameter ZT of certain materials by preparing them in the form of quantum-well superlattice structures was predicted. It has been also shown that CdTe nano-clusters embedded in PbTe lead to considerable changes of the derivative of the carrier density of states at the Fermi level and can influence the thermoelectrical
properties of the material.\cite{bukala2011} These theoretical results together with the recently reported fabrication of CdTe quantum dots in a PbTe matrix \cite{szot2011} open doors for using PbTe-CdTe structures to enhance the performance of thermoelectric devices.

Despite the mentioned above difficulties, bulk Pb$_{1-x}$Cd$_{x}$Te solid solutions in the form of polycrystalline samples were obtained by both, the Bridgman technique \cite{Ref.7} and by a rapid quenching followed by annealing.\cite{Ref.2,Ref.3,Ref.8} Recently, high quality single Pb$_{1-x}$Cd$_{x}$Te crystals with \emph{x} as high as 0.11 were obtained\cite{Ref.9} by self-selecting vapor growth method.\cite{Ref.31} These efforts were motivated by one more advantage in using PbTe as a base for forming ternary alloys, i.e., by the fact that in these materials the relative contributions of light and heavy holes, thus the electrical and optical properties of the system, can be tuned by changing the composition or temperature. Indeed, it was shown that while the energy gap of PbTe increases with Cd content, the energy separation between the light and heavy hole valence bands is considerable reduced.\cite{Ref.2,Ref.3}
Similar behavior was observed also in Pb$_{1-x}$Mn$_{x}$Te crystals.  In contrast to PbTe-CdTe system, the solid solution of PbTe and MnTe leads relatively easy to Pb$_{1-x}$Mn$_{x}$Te single crystals with $x$ up to $0.10$. The experimental studies of PbMnTe suggest that in this material adding Mn ions to PbTe changes also the relative positions of different valence band maxima, offering a possibility of improving the thermoelectric properties. In the p-type Pb$_{1-x}$Mn$_{x}$Te crystals it was shown that at room temperature and for constant carrier concentration ($p=2\times10^{18}cm^{-3}$) the thermoelectric power increases rapidly with the increase of Mn content, thus improving the thermoelectric figure of merit parameter Z.\cite{Ref.4}

In this paper we present a systematic study of the structural and electronic properties of Pb$_{1-x}$Cd$_{x}$Te and Pb$_{1-x}$Mn$_{x}$Te ternary alloys, for which either \emph{ab initio} or tight-binding methods were used, when appropriate. The first principle calculations of the stability of mixed crystals are presented in Sec. II. In Sec.~III, improved tight-binding description of the valence and conduction bands of PbTe, rock-salt CdTe and MnTe bulk crystals and the results obtained for the band structure of their solid solutions are shown. Section IV contains our conclusions, in particular the predicted, at using the obtained band structures, thermoelectric properties of the mixed crystals are discussed.

\section{Stability of the mixed-crystals}

In Pb$_{1-x}$Cd$_{x}$Te mixed crystals the transition from a ten-electron (\emph{x}=0) to an eight-electron system (\emph{x}=1) occurs, which is accompanied by a change in the crystal structure from rock-salt (RS) to zinc-blende (ZB). The fundamental question to be asked for these crystals is the \emph{x} value of this structure change. To answer this question, we analyze the stability of the crystal structure in Pb$_{1-x}$Cd$_{x}$Te alloys using \emph{ab initio} density functional theory (DFT) method. Moreover, using the same method, we determine the dependence of the lattice parameter on Cd and Mn content \emph{x} in Pb$_{1-x}$Cd$_{x}$Te and Pb$_{1-x}$Mn$_{x}$Te crystals, respectively. The calculations are performed within the Vienna \emph{ab initio} simulation package (VASP). \cite{Ref.12,Ref.13} For the atomic cores the projector augmented wave (PAW) pseudopotentials \cite{Ref.14} are used. The exchange correlation energy is calculated using the local density approximation (LDA). The atomic coordinates are relaxed with a conjugate gradient technique. The criterion that the maximum force is smaller than 0.01 eV/{\AA} is used to determine equilibrium configurations. In all the calculations, the energy cutoff is set to 16 Ry for the plane-wave basis, which is sufficient to obtain converged structural properties. Since the impact of nonscalar relativistic effects on the structural features is negligible \cite{Ref.15,Ref.16} we do not take these effects into account. Properties of Pb$_{1-x}$Cd$_{x}$Te and Pb$_{1-x}$Mn$_{x}$Te systems are calculated with ($2\times2\times2$) simple cubic supercells containing 64 atoms. The Brillouin zone integrations are performed using $4\times4\times4$ Monkhorst-Pack k-points meshes. All the atomic positions and the volume of the supercells are calculated with relaxation and re-bonding allowed. Our study is carried out for zero pressure and zero temperature.

 To study the stability of the crystal structure of Pb$_{1-x}$Cd$_{x}$Te alloys we calculate the total energy of RS PbTe and ZB CdTe supercells, in which we exchange successively the Pb(Cd) cations by Cd(Pb) ions, respectively. Thus, in the supercell of RS structure for Cd content $x = 0$ we have a PbTe crystal while for $x = 1$ we obtain a RS CdTe crystal. On the other hand, considering the ZB supercell, for $y = (1 - x) = 0$ we obtain the CdTe crystal and for $y = 1$ a hypothetical ZB PbTe crystal. One Cd atom in the RS PbTe supercell corresponds approximately to the Cd content $x= 0.03$ in RS Pb$_{1-x}$Cd$_{x}$Te alloy (similarly one Pb atom in the ZB CdTe supercell leads to ZB Pb$_{0.03}$Cd$_{0.97}$Te).

 %---------------------------------Energy difference PbTe-CdTe----------------------------------
\begin{figure}[!h]
\centering
\includegraphics[angle=-90,width=.42\textwidth]{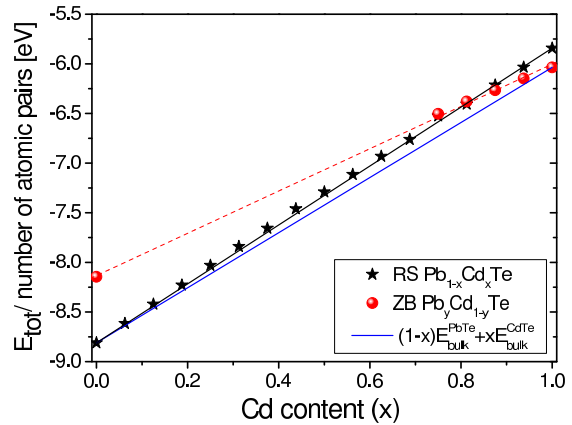}
\caption{\label{PbCdTe_energy}(Color online) Dependence of the total energy of RS Pb$_{1-x}$Cd$_{x}$Te (black stars) and the ZB Pb$_{y}$Cd$_{1-y}$Te (red dots) on the Cd content. The total energy is normalized to the number of cation-anion pairs. This energy is compared with an appropriate for the Cd content \emph{x} sum of the energies of two separate, RS PbTe and ZB CdTe, phases (blue line).}
\end{figure}
%---------------------------------------------------------------------------------------

Fig.~\ref{PbCdTe_energy} presents the dependence of the total energy (taken per atomic pair) on the Cd content for RS and ZB Pb$_{1-x}$Cd$_{x}$Te crystals. First, we see that the energy difference between RS and ZB structure is much bigger for $x = 0$ than for $x = 1$ (the energy of ZB PbTe is higher than that of RS PbTe by $\sim$ 0.666 eV/atomic pair, while for CdTe the RS structure leads to higher energy than ZB by $\sim$ 0.194 eV/atomic pair). We also observe that the RS structure of PbTe  is preserved in the calculations for Pb$_{1-x}$Cd$_{x}$Te mixed crystals  practically in the whole range of Cd concentrations. In contrast, adding Pb to CdTe destroys very quickly the ZB structure -- after relaxation the structure is maintained only  for Pb content up to ca $y = 1-x = 0.25$ (in our PbCdTe supercell with 32 cations, this corresponds to 8 atoms of Pb and 24 atoms of Cd). Therefore, in Fig.~\ref{PbCdTe_energy} only the points (red dots) which correspond to the preserved ZB structure of Cd$_{1-y}$Pb$_{y}$Te crystals are shown. These results are consistent with the observation that it is easier to obtain Pb$_{1-x}$Cd$_{x}$Te crystals than Cd$_{1-y}$Pb$_{y}$Te. Indeed, while there are no reports on CdTe highly doped by Pb, the successful growth of Pb$_{1-x}$Cd$_{x}$Te monocrystals with Cd content up to \emph{x} = 0.11 was reported.\cite{Ref.9}

We want to check for which Cd concentration the transition from RS to ZB structure should occur. As one can notice in Fig.~\ref{PbCdTe_energy}, the total energy of the mixed crystal for RS structure is lower  than for ZB for nearly whole range of Cd concentrations, i.e., up to  \emph{x} $\simeq$ 0.8. This means that virtual Pb$_{1-x}$Cd$_{x}$Te crystals with such high Cd concentrations would still have RS structure. However, for any \emph{x} value this energy is slightly higher than the total energy of separate phases (blue line in the Fig. \ref{PbCdTe_energy}) which we calculate as: $(1-x)E^{PbTe}_{bulk}+ xE^{CdTe}_{bulk}$. Still, for \emph{x} up to $\sim$~0.25 the energy difference between the line representing the energy of separated phases and energy of the mixed crystal in RS structure is lower than $k_BT$ in the growth conditions ($\sim$ 0.1 eV). This result may denote that the highest Cd content in samples presented in Ref.~\onlinecite{Ref.9} is close to a fundamental solubility limit.

\begin{figure}[!h]
\centering
\includegraphics[angle=-90,width=.43\textwidth]{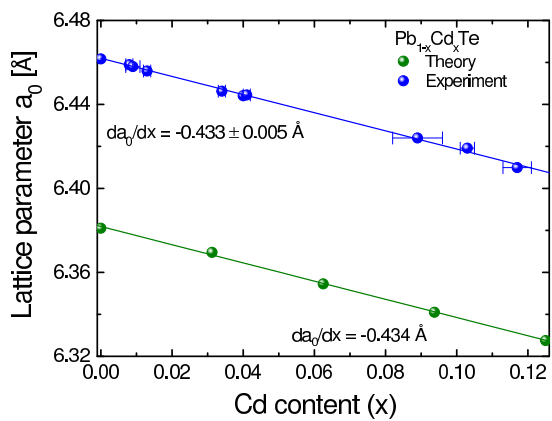}
\caption{\label{PbCdTe_lattice} (Color online) Lattice parameter of Pb$_{1-x}$Cd$_{x}$Te mixed crystal as a function of the Cd content \emph{x}. Green and blue dots denote the calculated and experimental values, respectively.}
\end{figure}

\begin{figure}[!h]
\centering
\includegraphics[angle=-90,width=.43\textwidth]{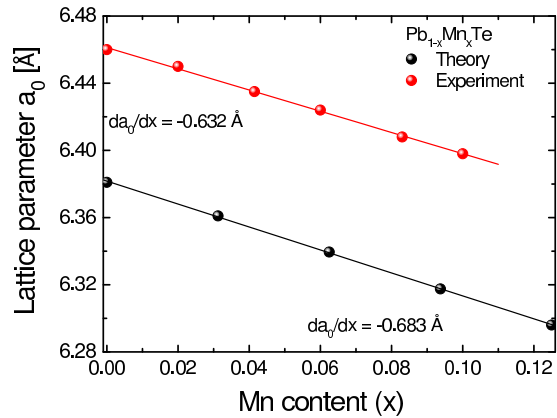}
\caption{\label{PbMnTe_lattice} (Color online) Dependence of the lattice parameter of Pb$_{1-x}$Mn$_{x}$Te on the Mn content \emph{x}. Black and red dots denote the calculated and experimental values, respectively.}
\end{figure}

In the next step we study the dependence of the lattice parameters \emph{$a_{0}$} of Pb$_{1-x}$Cd$_{x}$Te and Pb$_{1-x}$Mn$_{x}$Te crystals on the composition \emph{x}. For this purpose we consider a PbTe supercell in its stable RS crystal structure, in which we successively replace the Pb atoms by either Cd or Mn ions. The dependence of the calculated lattice parameter of Pb$_{1-x}$Cd$_{x}$Te on \emph{x} is presented in Fig.~\ref{PbCdTe_lattice} and for Pb$_{1-x}$Mn$_{x}$Te alloys in Fig.~\ref{PbMnTe_lattice}. We see that the relation between \emph{$a_{0}$} and $\emph{x} \leq 0.12$ for both mixed crystals is linear. The lattice parameters diminish with \emph{x} like $da_{0}/dx = -0.434$~\AA\ (Pb$_{1-x}$Cd$_{x}$Te) and $da_{0}/dx = -0.683$~\AA\ (Pb$_{1-x}$Mn$_{x}$Te). It is well known that the DFT calculations underestimate the lattice constants -- here the absolute values of the calculated lattice parameters are for both materials about $0.1$~\AA\ lower than the measured. Still, the obtained theoretically rate of decrease of the lattice parameter with \emph{x} for both crystals is in very good agreement with experimental data (compare Ref.~\onlinecite{Ref.9} for Pb$_{1-x}$Cd$_{x}$Te and Ref.~\onlinecite{Ref.32} for Pb$_{1-x}$Mn$_{x}$Te).
%---------------------------------Lattice constant PbCdTe PbMnTe----------------------------------

\section{Band structure of the constituent materials and their alloys}

  As shown above, the DFT calculations give much too low values of the lattice parameters. It is also well known that both, the local density and generalized gradient (LDA and GGA) approximations to the DFT method, widely used in the band structure calculations, underestimate the fundamental band gaps. In the case of narrow gap semiconductors, like PbTe and other lead chalcogenides, this leads even to a change of the sign of the gap, i.e., to an erroneous "inverted" order of bands at the L point of the Brillouin zone. Several attempts to overcome this drawback of the \emph{ab initio} calculations, by using GW method \cite{Svane}, hybrid functionals \cite{Heyd} or by simple rigid upward shift of the conduction band\cite{Wei}, were reported in the literature. Recently, it was shown\cite{Lusakowski} that correct band structure of PbTe and correct changes of the electronic structure of PbMnTe with the concentration of Mn ions and pressure can be obtained within DFT by reducing the spin-orbit strength for Pb $6p$ electrons by approximately 40$\%$.

Here virtual-crystal approximation and tight-binding method are exploited in the analysis of the electronic structure of Pb$_{1-x}$Cd$_{x}$Te and Pb$_{1-x}$Mn$_{x}$Te  alloys. Tight-binding methods have proven to be very useful in studying the electronic properties of solids. In the empirical tight-binding Hamiltonian, the matrix elements between orbitals centered on different sites are treated as parameters, which are adjusted to any known, often experimental values.
The band structures of the IV-VI semiconductor compounds obtained by tight-binding method\cite{Ref.17,Ref.18} are widely used to explain the observed phenomena in these materials.
To describe the band structure of the PbTe-CdTe and PbTe-MnTe solid solutions we need the tight-binding parametrization for all the constituent compounds, i.e., PbTe, CdTe and MnTe. We recall that the typical structure of Pb$_{1-x}$Cd$_{x}$Te and Pb$_{1-x}$Mn$_{x}$Te mixed crystals is the RS phase. Therefore, to examine the properties of these systems, we need parameters describing not only lead telluride, but also cadmium telluride and manganese telluride semiconductors in RS structure. While PbTe crystalizes in RS structure and RS MnTe is also common, CdTe changes its structure from ZB to RS only under high pressure. Therefore, the tight-binding parametrization for the two former materials can be found in literature, but to our best knowledge, there is no tight-binding parameters available for RS CdTe.

\subsection{PbTe}

\begin{figure}[!h]
\centering
\includegraphics[angle=-90,width=.43\textwidth]{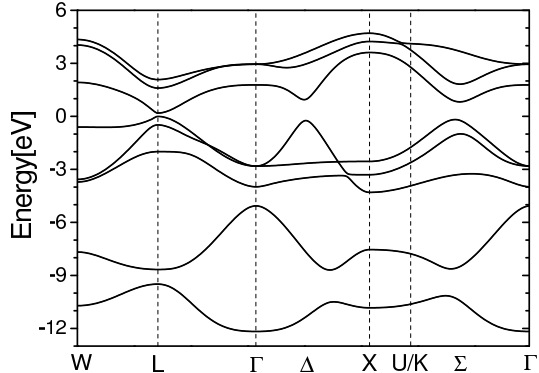}
\caption{\label{PbTe} The calculated within our tight-binding model band structure of PbTe along the symmetry lines of the Brillouin zone.}
\end{figure}

Let us start from the band structure of PbTe.~A careful analysis of the tight-binding parameters available in the literature \cite{Ref.17,Ref.18} shows that although they recover correctly the band structure, they do not lead to effective masses, which are determined experimentally. Thus, in order to fit our tight-binding model to all existing experimental results, we have performed a new tight-binding parametrization for the PbTe crystal. To describe PbTe material we use the $sp^{3}$ atomic orbitals, with the spin-orbit coupling included. In our model we consider the nearest-neighbor cation-anion as well as next nearest anion-anion and cation-cation inter-atomic couplings. We use the experimentally determined energy gap in low temperatures, i.e., $E_g = 0.19$ eV, \cite{Dalven} and assume that at T = 0K the second valence band maximum along the $\Sigma$-line is located about 0.17 eV below the top of the valence band at the L point, as suggested in Ref.~\onlinecite{Ref.20}. Another experimental input to our fitting comes from Ref.~\onlinecite{Ref.21}, where the values of the longitudinal and perpendicular PbTe effective masses at the L point of the Brillouin zone are given. Spin-orbit parameters are matched to the atomic values for Pb and Te atoms, which are equal 1.273 eV and 0.840 eV, respectively.\cite{Ref.22} In our fitting procedure we allow for an adjustment of the latter parameters, but we keep the ratio of cation to anion spin-orbit coupling parameters  equal to the Pb/Te atomic values rate, i.e., to 1.51. Our final values for spin-orbit parameters differ from the atomic spin-orbit couplings by less than $10\%$ (compare: Table \ref{Tab.1}). In Table \ref{Tab.1} the obtained in this work tight-binding parameters for all, PbTe, RS MnTe and RS CdTe, materials are shown in the standard Slater-Koster notation. \cite{Ref.23}
%---------------------------------Parameters----------------------------------
\begin{table}[!h]
\caption{\label{Tab.1}Nearest- and next-nearest neighbors tight-binding parameters and the on-site energies for RS PbTe, CdTe and MnTe crystals. All energies are given in eV and the energy zero is always assumed at the top of the valence-band. The $\triangle$ denotes the spin-orbit coupling parameter.}
\centering
\begin{ruledtabular}
\begin{tabular}{lrrr}
Parameters (in eV)& PbTe & CdTe & MnTe\\\hline
$E_{s_c}$ & -8.6528 & -2.3095& 1.6517\\
$E_{p_c}$ &  1.2711 &  3.7778& 4.5995\\
$E_{d_c}$ &          & -8.0291& 0.1676\\
$E_{s_a}$ & -9.4379 & -9.7514& -9.5918\\
$E_{p_a}$ & -0.8324&  1.0687& 0.2887\\
\\
$s_{c}s_{a}\sigma$ & 0.3265 & 0.9943   & 0.9823\\
$s_{c}p_{a}\sigma$ & 0.0838 & 1.3711   & 1.9093\\
$p_{c}s_{a}\sigma$ & 0.2148 & 0.9478   & -0.0786\\
$p_{c}p_{a}\sigma$ & 1.6702  & 2.0861   & 2.5137\\
$p_{c}p_{a}\pi$    & -0.1149& -0.6885 & -0.2744\\
$d_{c}s_{a}\sigma$ &          & 0.2070  & 0.1959\\
$d_{c}p_{a}\sigma$ &          & 0.7366  & 0.5445\\
$d_{c}p_{a}\pi$    &          & -0.0709   & 0.5420\\
\\
$s_{c}s_{c}\sigma$ & -0.2444& -0.0951   & 0.0954\\
$s_{c}p_{c}\sigma$ &  0.4909&  0.2331 & -0.1853\\
$p_{c}s_{c}\sigma$ & -0.4909& -0.2331 & 0.1853\\
$p_{c}p_{c}\sigma$ & -0.0160& 0.5027  & -0.1469\\
$p_{c}p_{c}\pi$    & -0.1869 & 0.1106  & -0.1188\\
$d_{c}s_{c}\sigma$ &          & 0.0713   & -0.1645\\
$d_{c}p_{c}\sigma$ &          & -0.3165   &  0.3084\\
$d_{c}p_{c}\pi$    &          & 0.3207  & -0.1863\\
$d_{c}d_{c}\sigma$ &          & -0.1699   & -0.1402\\
$d_{c}d_{c}\pi$    &          & 0.0299   &  0.0244\\
\\
$s_{a}s_{a}\sigma$ &  0.3153 & 0.1633   & -0.0703\\
$s_{a}p_{a}\sigma$ & 0.3874 & 0.1037  & 0.0021\\
$p_{a}s_{a}\sigma$ &-0.3874 & -0.1037 & -0.0021\\
$p_{a}p_{a}\sigma$ & 0.2121 & -0.4561   & 0.3351\\
$p_{a}p_{a}\pi$    &-0.0467&  0.0737 & 0.0553\\
\\
$\Delta_{c}/3$     & 0.4692  & 0.2924  &      \\
$\Delta_{a}/3$     & 0.3109 & 0.4884  &      \\
\end{tabular}
\end{ruledtabular}
\end{table}
%-----------------------------------------------------------------------------------
The band structure of PbTe resulting from our tight-binding parametrization is presented in Fig.~\ref{PbTe}. As one can see in the Figure, our improved parametrization gives the correct value of the energy gap at the L point and the appropriate energy position of the second valence band maximum at the $\Sigma$. Moreover, in contrast to the earlier calculations, our model provides proper values of the longitudinal and perpendicular effective masses. In Table~\ref{Tab.2} the comparison of the effective masses obtained within our model and the other theoretical approaches with the experimental values is presented.

%---------------------------------Effective masses----------------------------------
\begin{table}[!h]
\caption{\label{Tab.2}Experimental and theoretical values of the effective masses of PbTe at the L point. $m^{*}_{l}$, $m^{*}_{t}$ denote the longitudinal and transversal effective masses of holes (\emph{h}) and electrons (\emph{e}), respectively.}
\begin{ruledtabular}
\begin{tabular}{l|l|l|l|l}
 & $m^{*}_{l_h}/m_{0}$ & $m^{*}_{t_h}/m_{0}$ & $m^{*}_{l_e}/m_{0}$ & $m^{*}_{t_e}/m_{0}$\\\hline
Calc.\cite{Ref.17} & 0.162 & 0.033 & 0.133 & 0.0281\\
Calc.\cite{Ref.18} & 0.0799 & 0.0133 & 0.0799 & 0.0107\\
This work & 0.294 & 0.0276 & 0.272 & 0.0241\\
Expt.\cite{Ref.21} & 0.31$\pm$0.05 & 0.022$\pm$0.03 & 0.24$\pm$0.05 & 0.024$\pm$0.03
\end{tabular}
\end{ruledtabular}
\end{table}
%---------------------------------------------------------------------------------------
%---------------------------------Band structure PbTe----------------------------------
%---------------------------------------------------------------------------------------

\subsection{CdTe in rock-salt structure}

In the second step we analyze the cadmium telluride in RS structure. To our knowledge, there are only few papers, which report obtaining RS CdTe by applying high pressure to the ZB CdTe crystals. In Refs.~\onlinecite{Ref.27,Ref.25} the optical studies of the band structure of this material are reported. The transition from ZB to the RS phase is observed at (3.8 $\pm 0.2$)~GPa as a dramatic decrease of the sample transmittance (the samples become virtually opaque between 3.9 GPa and
4.5 GPa). The measurements revealed that the change from ZB CdTe to RS CdTe results in a characteristic shift of the valence band maximum away from the $\Gamma$ point towards the L and K points, due to p--d hybridization effects. As a result, two types of indirect gaps occur. However, it is very difficult to determine experimentally the indirect band gaps of RS CdTe, because during the phase transition from tetrahedral to octahedral coordination the large number of defects and dislocations induced by the applied pressure form band-tails states.\cite{Ref.27}

The band structure of CdTe in RS structure was already determined by using DFT-LDA method. In the  calculation performed in Ref.~\onlinecite{Ref.25}, RS-CdTe turned out to be a semimetal, in which the conduction band minimum at the X point would be nearly 2.5 eV below the valence band maximum located at $\Sigma$, midway the $\Gamma$ -- K line. This calculation, as well as another one presented in Ref.~\onlinecite{Ref.24}, does not take into account the spin-orbit effects, which are important in the analysis of the electronic properties of CdTe. The calculations of band structure with spin-orbit coupling were performed, using linear muffin-tin orbital method, by Christensen and Christensen.\cite{Ref.26} The band structure obtained in the latter exhibits also a metallic character. Additionally, the band gap at the $\Gamma$ point, at the center of Brillouin zone, is very close to zero.
It should be emphasized that the authors of all the mentioned above results attribute the obtained semi-metallic character of RS-CdTe to the limitations of DFT-LDA methods only, i.e., to the fact that these calculations underestimate considerably the band gaps and can even give false overlaps between valence and conduction bands. Indeed, reflectance experiments in the mid-infrared do not indicate a metallic behavior of RS CdTe.\cite{Ref.25} G\"{u}der \emph{et al.} after a careful analysis of the experimental results obtained for RS-CdTe at different pressures made an assumption that RS-CdTe is a narrow gap semiconductor, with the energy gap of few hundreds of meV at the $\Gamma$ point.\cite{Ref.25} This assumption is consistent also with results of Ref.~\onlinecite{Ref.27}.

\begin{figure}[!h]
\centering
\includegraphics[angle=-90,width=.43\textwidth]{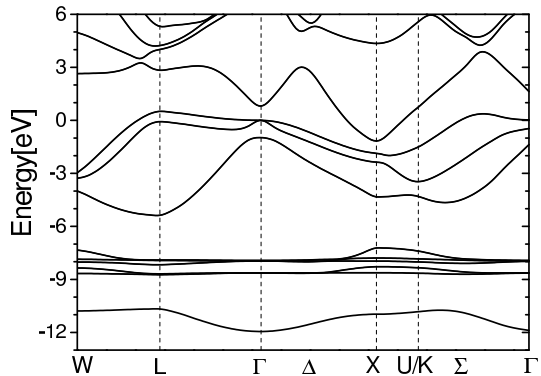}
\caption{\label{CdTe-RS} RS-CdTe band structure along the main symmetry lines of the Brillouin zone, as calculated by using DFT-LDA method within the VASP code.}
\end{figure}

As described above, there are neither exact experimental data nor a reliable theoretically calculated band structure, to which we can fit our tight-binding parameters for RS CdTe bulk crystal. In this situation we decided to try to obtain a better description of the band structure of RS CdTe in high symmetry directions of the fcc Brillouin zone by using the \emph{ab initio} procedure described in Sec. II. In this calculations the spin-orbit interactions are taken into account in the VASP code. The results are presented in Fig.~\ref{CdTe-RS}. As one can notice in Fig.~\ref{CdTe-RS}, we also obtain that the conduction band minimum of RS CdTe is located at the X point and is below the valence band maximum near the L point. The band structure is qualitatively similar to the structure presented in Ref.~\onlinecite{Ref.26}, but in our case the band gap at $\Gamma$ point equals {$\sim 0.805$ eV}.  Thus, the presented in Fig.~\ref{CdTe-RS} structure while repeating the drawbacks of the other \emph{ab initio} DFT structures, has the advantage of agreeing with the assumption of few hundreds of meV energy difference between the conduction and valence bands at the $\Gamma$ point.  The parameters for this material presented in Table \ref{Tab.1} are obtained by fitting the tight-binding band structure to the one presented in Fig.~\ref{CdTe-RS}.

%---------------------------------Band structure CdTe-RS----------------------------------
%---------------------------------------------------------------------------------------

\subsection{MnTe}

\begin{figure}[!h]
\centering
\includegraphics[angle=-90,width=.43\textwidth]{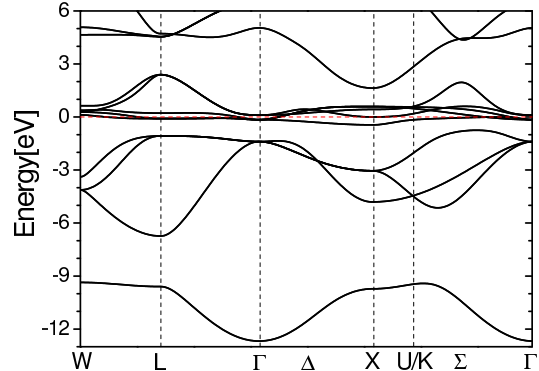}
\caption{\label{MnTe_RS} The calculated within tight-binding model (with nearest- and next-nearest-neighbors interactions taken into account) band structure of MnTe in RS structure. The red dashed line is the Fermi level. }
\end{figure}

The band structure of manganese salts, particularly MnTe, has been the subject of several different models. \cite{Ref.28,Ref.29,Ref.30} According to Allen \emph{et al.},\cite{Ref.29} the electronic structure of MnTe is determined by the combined effect of an exchange splitting of the Mn 3d states and of a strong hybridization of these states with the anionic p states. The hybridization tends to delocalize the d electrons and mediates the contribution of the d states to the structure of the valence and conduction bands. The impact of hybridization effects on the total band structure depends on the position of the d states. A full tight-binding description of the electronic structure of RS MnTe, which is based on the spin-fluctuation theory, was presented by Ma\u{s}ek \emph{et al}. \cite{Ref.30} The obtained band structure exhibits metallic character with the Fermi level fixed within the half-filled d band. The top of the valence band is at L and the bottom of the conduction band is at X. In this tight-binding calculations, however, the spin-orbit interactions were not included and only the interactions between the nearest-neighbors were taken into account. This is not consistent with our description of PbTe bulk crystal, where we consider not only the interactions between the nearest-neighbors but also next-nearest neighbor tight-binding integrals. For a proper description of Pb$_{1-x}$Mn$_{x}$Te mixed crystals, in particular for applying the virtual-crystal approximation, it is most reasonable to take into account the same number of neighbors in both constituent materials. Thus, we decided not to take for the tight-binding description of MnTe crystal the parametrization of  Ref.~\onlinecite{Ref.30}. Instead, we have considered a model with the nearest-neighbor cation-anion as well as next nearest anion-anion and cation-cation inter-atomic couplings, like done before for PbTe. Still, the model parameters were fitted to reproduce the band structure obtained by Ma\u{s}ek \emph{et al}.\cite{Ref.30}
The electronic structure of RS MnTe resulting from our model is shown in Fig.~\ref{MnTe_RS}. It should be noted, that in our tight-binding description of RS MnTe the spin-orbit splittings could not be included, because they were not present in the structure in Ref.~\onlinecite{Ref.30}, which we tried to reproduce. The $sp^{3}d^{5}$ {tight-binding} parameters for RS MnTe are presented in Table~\ref{Tab.1}.

%---------------------------------Band structure MnTe-RS----------------------------------
%---------------------------------------------------------------------------------------

\subsection{Pb$_{1-x}$Cd$_{x}$Te and Pb$_{1-x}$Mn$_{x}$Te mixed crystals}

\begin{figure}[!ht]
\centering
\includegraphics[angle=-90,width=.45\textwidth]{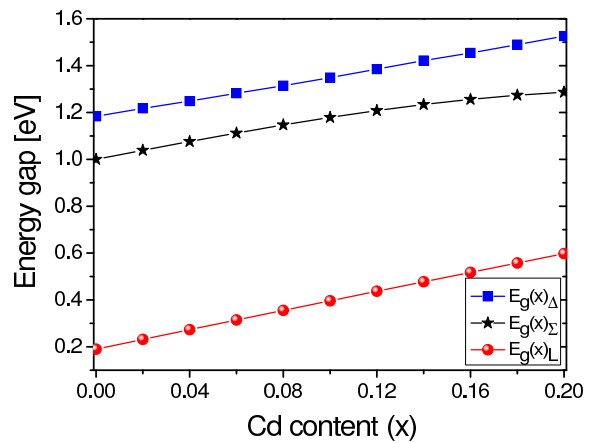}
\caption{\label{PbCdTe_Eg} (Color online) Dependence of the energy gaps in Pb$_{1-x}$Cd$_{x}$Te on the Cd content \emph{x}. The blue squares, black stars and red dots denote the energy gaps at $\Delta$, $\Sigma$ and L, respectively.}
\end{figure}

\begin{figure}[!h]
\centering
\includegraphics[angle=-90,width=.45\textwidth]{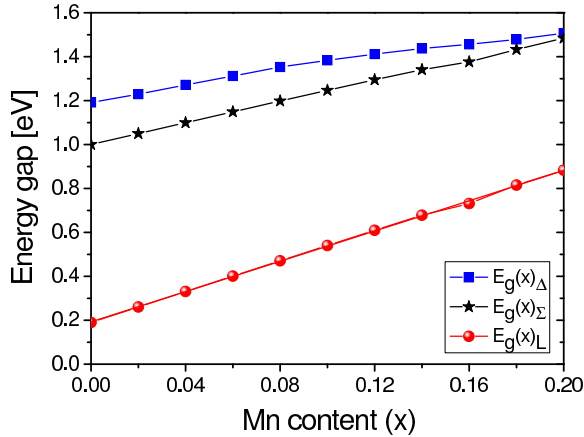}
\caption{\label{PbMnTe_Eg} (Color online) Dependence of the energy gaps in Pb$_{1-x}$Mn$_{x}$Te on the Mn content \emph{x}. The blue squares, black stars and red dots denote the energy gaps at $\Delta$, $\Sigma$ and L, respectively.}
\end{figure}

%-----------------------------------------------------------------------------------------------------
 To calculate the band structures of Pb$_{1-x}$Cd$_{x}$Te and Pb$_{1-x}$Mn$_{x}$Te alloys within the tight-binding approach, we apply the virtual crystal approximation, i.e., all the on-site energies and interaction integrals are assumed to be in $x$ part equal to the given parameter for CdTe or MnTe, and in the remaining $(1-x)$ part equal to that of PbTe. In Fig.~\ref{PbCdTe_Eg} the dependence of the energy gap of Pb$_{1-x}$Cd$_{x}$Te crystals in L, $\Sigma$ and $\Delta$ points of Brillouin zone on the Cd content, up to $x = 0.2$, is presented. The calculated band gap in the L point is almost a linear function of composition and increases with \emph{x} like $dE_{g}/dx\approx 2$~eV. It should be noted that this is ca 30\% faster increase than that suggested by the experimental results of Ref.~\onlinecite{Ref.9}. Fig.~\ref{PbMnTe_Eg} presents the results of similar calculation for Pb$_{1-x}$Mn$_{x}$Te alloy for Mn concentrations also up to $x = 0.2$. As shown in the Figure, the energy gap in the L point increases with the Mn content ca 35 meV/at \%. The obtained increase of the band gap of Pb$_{1-x}$Mn$_{x}$Te with the Mn content is also slightly more rapid than the value 25 meV/(at \% of Mn) estimated from the experimental data in Ref.~\onlinecite{Ref.4}. On the other hand, however, the DFT calculations presented in Ref.~\onlinecite{Lusakowski} lead to probably too small slope of 15 meV/(at \% of Mn).

\begin{figure}[!h]
\centering
\includegraphics[angle=-90,width=.45\textwidth]{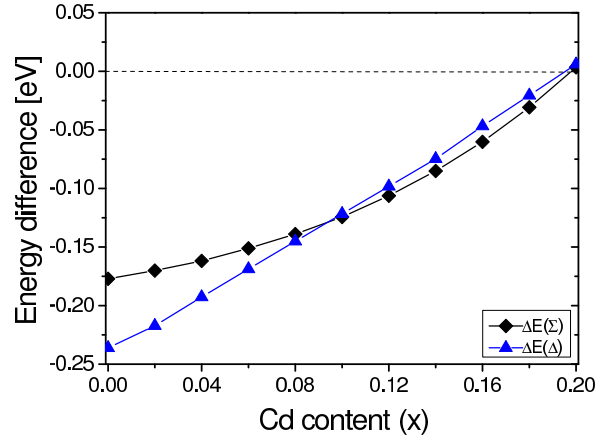}
\caption{\label{PbCdTe_DE} (Color online) Energy differences between the valence band maximum at L and $\Sigma$ (black rhombuses) and between L and $\Delta$ (blue triangles) on the Cd concentration \emph{x} in Pb$_{1-x}$Cd$_{x}$Te.}
\end{figure}

\begin{figure}[!h]
\centering
\includegraphics[angle=-90,width=.45\textwidth]{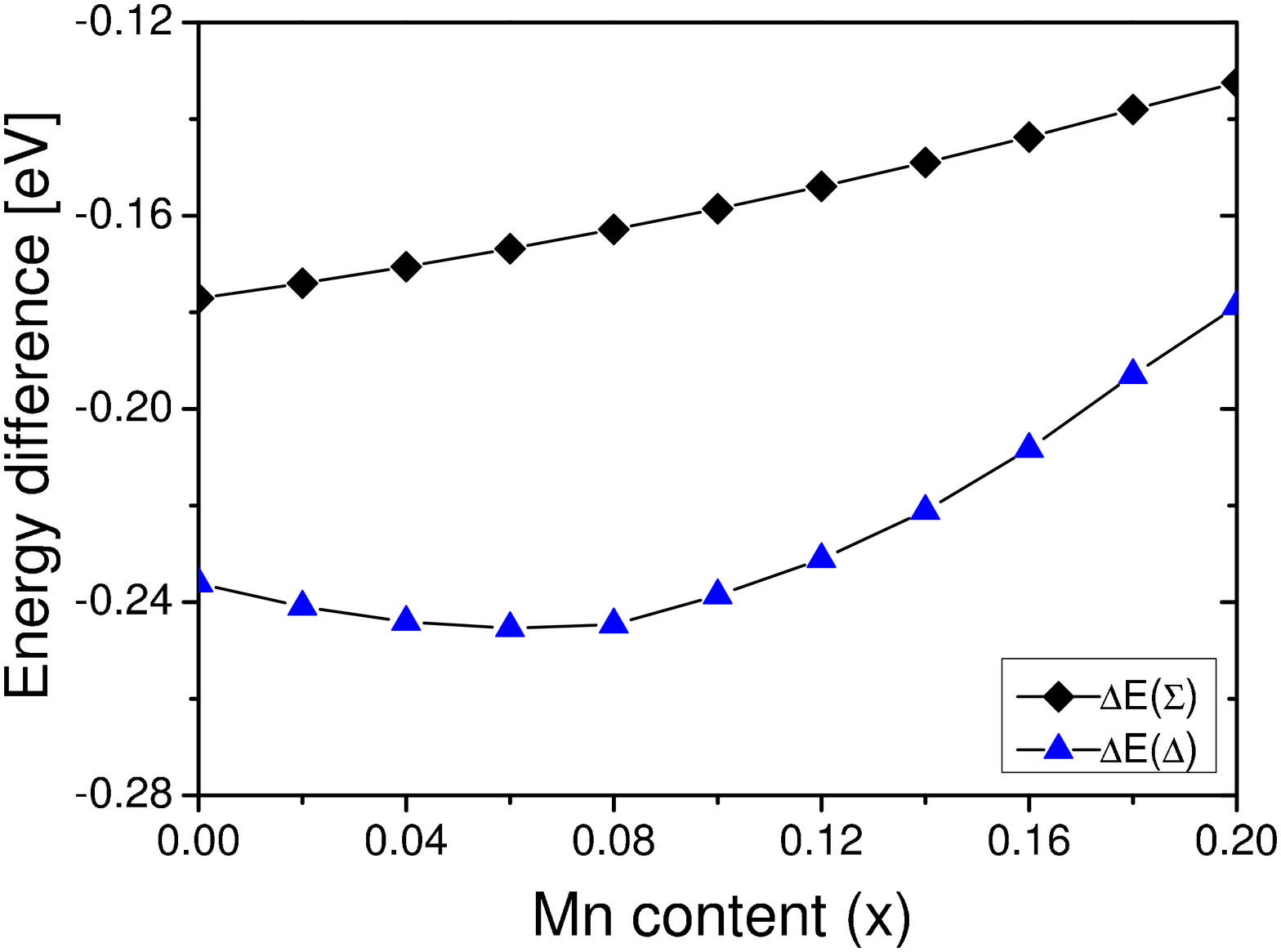}
\caption{\label{PbMnTe_DE} (Color online) Energy differences between the valence band maximum at L and $\Sigma$ (black rhombuses) and between L and $\Delta$ (blue triangles) on the Mn concentration \emph{x}in Pb$_{1-x}$Mn$_{x}$Te.}
\end{figure}

It should be recalled here that our study of PbTe-based mixed crystals was mainly motivated by the idea that in these materials a change of relative position of the heavy and light hole valence bands can lead to higher thermoelectric power  as compared to the p-type PbTe. It can lead to an increased ZT because the total thermoelectric power factor for the material derives from the contributions from all extrema.\cite{Hicks, Larson} Also in Ref.~\onlinecite{Ref.4} the strong increase of thermoelectric power found in Pb$_{1-x}$Mn$_{x}$Te was explained by assuming a change in sign of the separation energy between the band extremum of light holes at the L point and the band of heavy holes at $\Sigma$. A similar behavior can be expected in the Pb$_{1-x}$Cd$_{x}$Te crystals. Indeed, in Figs~\ref{PbCdTe_Eg} and \ref{PbMnTe_Eg} we observe that the energy distance between the valence and conduction bands in the other extrema (at $\Delta$ and $\Sigma$) increases with adding Cd or Mn much slower than that in the L maximum. To study this behavior more carefully, we determine the energy differences between the valence band maximum at the L point and another maximum at the $\Sigma$, as well as between L and $\Delta$, i.e., we determine the so called "side energy gaps". The obtained side gaps as a function of the composition $x$ for Pb$_{1-x}$Cd$_{x}$Te  are presented in Fig.~\ref{PbCdTe_DE} and for Pb$_{1-x}$Mn$_{x}$Te in Fig.~\ref{PbMnTe_DE}.

First we notice that according to our calculations in Pb$_{1-x}$Mn$_{x}$Te, although the side energy gap $\Sigma$ - L diminishes with $x$, up to x=0.2 the maximum at $\Sigma$ is lower in energy than the maximum at the L point of the Brillouin zone. Thus, this is a slower decrease than that suggested in Refs~\onlinecite{Ref.4} and \onlinecite{Lusakowski} (within the \emph{ab initio} calculations with reduced spin-orbit interactions, presented in \onlinecite{Lusakowski}, it was obtained that for Mn concentrations higher than $x\approx 0.09$ the $\Sigma$ appears above the L maximum). For Pb$_{1-x}$Cd$_{x}$Te we obtain that for $x\geq0.1$ the band maximum at $\Delta$ appears above the top of the valence band at $\Sigma$ and that both side energy gaps should change sign for $x\cong 0.2$, as shown in Fig.~\ref{PbCdTe_DE}. Despite these quantitative differences, our results for both materials, Pb$_{1-x}$Mn$_{x}$Te and Pb$_{1-x}$Cd$_{x}$Te, qualitatively agree with the idea that in these mixed crystals higher $x$ enhances the role of the heavy holes from the vicinity of the other maxima of valence band. This is especially valid for higher concentrations of the p-type carriers in the samples.

\section{Summary and Discussion}

 Our analysis of the stability of Pb$_{1-x}$Cd$_{x}$Te, performed by \emph{ab initio} methods, shows that up to $x \simeq 0.8$ the total energy of the ternary alloy is lower for the RS structure. This result is consistent with the observation that it is easier to obtain rock-salt Pb$_{1-x}$Cd$_{x}$Te crystals with considerable amount of Cd than zinc-blende CdTe crystals doped with Pb. The lattice parameters of Pb$_{1-x}$Cd$_{x}$Te and (Pb$_{1-x}$Mn$_{x}$Te) mixed crystals decrease with Cd (Mn) content \emph{x}.  The lattice parameters $a_{0}$ diminish with \emph{x}$\leq 0.12$ like $da_{0}/dx \ = -0.434$~\AA\ (Pb$_{1-x}$Cd$_{x}$Te) and $da_{0}/dx \ = -0.683$~\AA\ (Pb$_{1-x}$Mn$_{x}$Te), in good agreement with the experimental findings.

 The energy structures of Pb$_{1-x}$Cd$_{x}$Te and Pb$_{1-x}$Mn$_{x}$Te alloys have been analyzed using tight-binding description of constituent materials and virtual crystal approximation.~For this purpose we have performed a tight-binding parametrization of rock-salt PbTe, CdTe and MnTe. In contrast to previous approaches, our tight-binding parameters of lead telluride lead not only to correct overall band structure of PbTe in the whole Brillouin zone but also accurately reproduce the experimental bulk effective masses. Due to the lack of exact experimental data for RS CdTe, the tight-binding description of this material is based on the results of DFT calculations - still, our model recovers the predicted band gap in the center of the Brillouin zone.  An increase of the L-point band gap with \emph{x} has been obtained for both studied ternary alloys. The calculated band gaps in the L maximum are almost linear functions of the composition \emph{x} and compare well with the experimental results. We have calculated also the energy differences between the valence band maximum at the L point and another maximum at $\Sigma$, as well as between L and $\Delta$. In both studied materials the side energy gaps $\Sigma$ - L and $\Delta$ - L diminish with $x$. Thus,  for higher concentrations of the p-type carriers in the samples an enhanced contribution of the heavy holes from the valence band in the vicinity of the other valence band extrema is predicted. To discuss how adding Cd or Mn to PbTe changes the thermoelectric properties of the material, we have calculated the Seebeck coefficient (S) in both ternary alloys, using above described band structures of the mixed crystals.

As a first approximation we consider Seebeck coefficient within Mahan-Sofo theory, which is given by the simplified Mott expression:\cite{Mott}

\begin{equation}
    S=\frac{\pi^2}{3}\frac{k^2_B}{q}T\bigg\{\frac{1}{D}\frac{dD(E)}{dE}+\frac{1}{\mu}\frac{d\mu(E)}{dE}\bigg\}_{E=E_F}
\end{equation}
where $E_F$ is the Fermi energy, $q$ is the carrier charge; $D(E)$ and $\mu(E)$ are the density of states and the mobility, both energy dependent.

\begin{figure}[!ht]
\centering
\includegraphics[angle=-90,width=.42\textwidth]{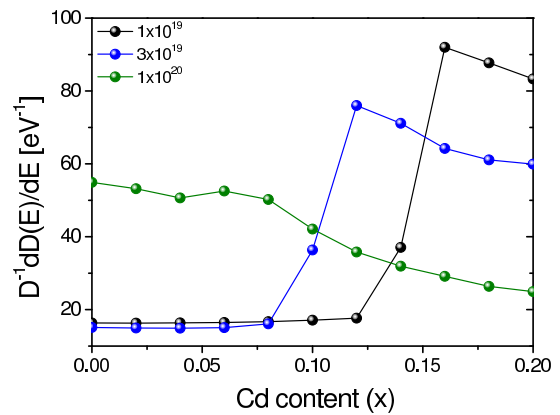}
\caption{\label{PbCdTe_density}(Color online) The values of the $D^{-1}dD(E)/dE$ expression at the Fermi level as a function of Cd content for $p=1\times 10^{19} cm^{-3}$ (black line), $p=3\times 10^{19} cm^{-3}$ (blue line) and $p=1\times 10^{20} cm^{-3}$ (green line).}
\end{figure}

In Fig.~\ref{PbCdTe_density} we show how the first term in the Mott equation (which describes the role played by the changes of the density of states in the thermoelectric power) depends on the content of Cd in Pb$_{1-x}$Cd$_{x}$Te. The calculations have been performed for three different hole concentrations: $p=1\times 10^{19} cm^{-3}$, $p=3\times 10^{19} cm^{-3}$ and $p=1\times 10^{20} cm^{-3}$, i.e., different positions of the Fermi level in the valence band. As one can see in the Figure, the $D^{-1}dD(E)/dE$ value increases dramatically whenever the top of the heavy hole band at the $\Delta$ point reaches the Fermi level. For $p=1\times 10^{19} cm^{-3}$ this happens for $x\approx 0.16$ and for $p=3\times 10^{19} cm^{-3}$ for $x\approx 0.12$. For the highest, $p=1\times 10^{20} cm^{-3}$, hole concentration the heavy hole band with the top at $\Sigma$ contributes to $D^{-1}dD(E)/dE$ value at the  Fermi level even for $x=0$. For this hole concentration the small increase at $x=0.06$ is attributed to the onset of the contribution from the $\Delta$ maximum (compare Figs \ref{PbCdTe_density} and \ref{PbCdTe_DE}). Here it is worth to recall that  the
concept of carrier pocket engineering to produce convergence of symmetrically inequivalent bands has been suggested first for low-dimensional thermoelectric nanostructures\cite{Mildred} and then extended to bulk materials.\cite{Peinature} In these papers it has been suggested that the convergence of many charge carrying valleys has an effect of producing large m* without explicitly reducing the mobility $\mu$ and that a large valley degeneracy should improve thermoelectric properties of the materials.  In Ref. \onlinecite{Peinature} this effect was shown in PbTe$_{1-x}$Se$_{x}$, where the L and $\Sigma$ valence bands can be converged, giving an increased valley degeneracy of 16. We note that according to our calculations, in Pb$_{1-x}$Cd$_{x}$Te, under proper conditions, the valley degeneracy as high as 22 can be achieved, due to additional contribution from the secondary valence band ($\Delta$ -- with the degeneracy equal to 6), which in this material is also very close to the L and $\Sigma$ bands.
%---------------------------------PbCdTe-density--------------------------------
%---------------------------------------------------------------------------------------

The presented above Seebeck coefficient enhancement through density of states modification is a
promising route,\cite{Heremans, Heremans2012}  but this approach risks the reduction of carrier mobility.\cite{Peinature}
The optimal electronic performance of a thermoelectric semiconductor depends strongly also on the weighted mobility.\cite{Mahan}
To study Seebeck coefficient more precisely, in particular to take into account the mobility term, we have followed the scheme proposed in the paper of Madsen and Singh,\cite{Madsen} i.e., we have performed the calculations using Boltzmann transport theory \cite{Allen, Ziman, Hurd} and constant relaxation time approximation.\cite{Nag} The latter approximation is based on an assumption that the scattering time determining the electrical conductivity does not vary strongly with energy.
The advantage of using constant relaxation time is that the thermoelectric power can be directly calculated from the band structure as a function of carrier concentration and temperature, with no adjustable parameters. In Ref.~\onlinecite{Madsen} the thermoelectric transport tensors (i.e., the electrical conductivity $\sigma(T, \mu)$ and Seebeck coefficient S(T, $\mu$)) are defined by the following expressions:
\begin{equation}
    S_{ij}=(\sigma^{-1})_{\alpha i}\nu_{\alpha j}
\end{equation}
where
\begin{equation}
    \sigma_{\alpha \beta}(T,\mu)=\frac{1}{\Omega}\int\sigma_{\alpha \beta}(\varepsilon)[-\frac{\partial f_\mu(T,\varepsilon)}{\partial\varepsilon}]d\varepsilon
\end{equation}
\begin{equation}
    \nu_{\alpha \beta}(T,\mu)=\frac{1}{eT \Omega}\int\sigma_{\alpha \beta}(\varepsilon)(\varepsilon-\mu)[-\frac{\partial f_\mu(T,\varepsilon)}{\partial\varepsilon}]d\varepsilon
\end{equation}
\noindent
and the transport distribution function tensor $\sigma_{\alpha\beta}(\varepsilon)$ is given by
\begin{equation}
    \sigma_{\alpha\beta}(\varepsilon)=\frac{e^{2}}{N}\sum_{i,\textbf{k}}\tau\upsilon_{\alpha}(i,\textbf{k})\upsilon_{\beta}(j,\textbf{k})\delta(\varepsilon-\varepsilon_{i,\textbf{k}})
\end{equation}
\noindent
where $\varepsilon_{i,\textbf{k}}$ is band energy, $\upsilon_{\alpha}(i,\textbf{k})$ is the band velocity
($\partial \varepsilon_{i,\textbf{k}}/\partial \textbf{k}_{\alpha}$); $\tau$, $\mu$, and $\emph{f}_{\mu}$ are the relaxation time, chemical potential, and Fermi-Dirac distribution function, respectively. More details about this model can be found in Ref. \onlinecite{Madsen}.

\begin{figure}[!ht]
\centering
\includegraphics[angle=-90,width=.42\textwidth]{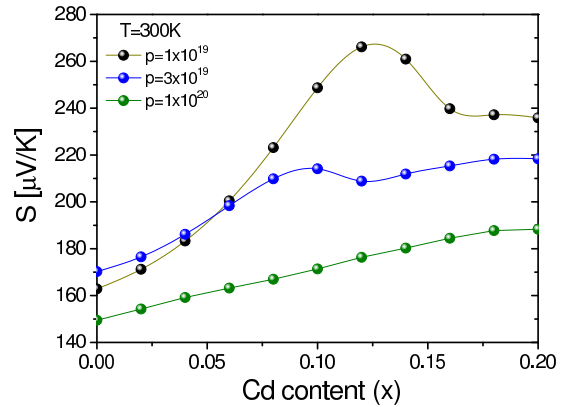}
\caption{\label{PbCdTe_seebeck}(Color online) The dependence of the calculated thermoelectric power of p-type Pb$_{1-x}$Cd$_{x}$Te crystals on the Cd content. The black line corresponds to $p=1\times 10^{19} cm^{-3}$, blue line to $p=3\times 10^{19} cm^{-3}$ and green line to $p=1\times 10^{20} cm^{-3}$.}
\end{figure}

\begin{figure}[!ht]
\centering
\includegraphics[angle=-90,width=.42\textwidth]{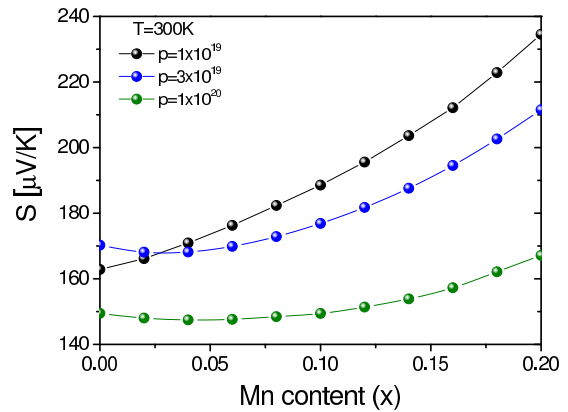}
\caption{\label{PbMnTe_seebeck}(Color online) The dependence of the calculated thermoelectric power of p-type Pb$_{1-x}$Mn$_{x}$Te crystals on the Mn content.The black line corresponds to $p=1\times 10^{19} cm^{-3}$, blue line to $p=3\times 10^{19} cm^{-3}$ and green line to $p=1\times 10^{20} cm^{-3}$.}
\end{figure}

The dependence of the thermoelectric power in p-type Pb$_{1-x}$X$_{x}$Te mixed crystals (X denotes either Cd or Mn) on the composition \emph{x} was calculated for the same three carrier concentrations ($p=1\times 10^{19} cm^{-3}$, $p=3\times 10^{19} cm^{-3}$ and $p=1\times 10^{20} cm^{-3}$) and room temperature 300K. Eigenvalues of $200\times200\times200$ k-points mesh were calculated by tight-binding method. The obtained results for Pb$_{1-x}$Cd$_{x}$Te are shown in Fig.~\ref{PbCdTe_seebeck} and for Pb$_{1-x}$Mn$_{x}$Te in Fig.~\ref{PbMnTe_seebeck}. As one can see in the Figures, in both materials the Seebeck coefficient increases with increasing Cd (Mn) content. The  thermoelectric  power  for  p-type  alloys with  high  carrier  density  differs  appreciably from the corresponding values for PbTe.  It should be noticed that this happens not only because  of the greater  significance of heavy-mass hole conduction, which results from  the  smaller  energy  separation  between  light and heavy hole valence bands. The increase results also from the more parabolic nature of the light-mass valence band (a consequence of the larger direct energy gap), as described already by Rogers and Crocker many years ago.\cite{Rogers} This is especially seen in Fig.~\ref{PbMnTe_seebeck}, because according to our models in  Pb$_{1-x}$Mn$_{x}$Te the admixture of the heavy holes contribution is not as strong as expected and it does not play a considerable role for the Mn content shown in the Figure. Still, according to our calculations, adding ca 10\% of Mn to PbTe should enhance the thermoelectric power by ca 15\%. In PbTe with the same amount of Cd a much higher thermoelectric power should be observed. The results presented in Figs \ref{PbMnTe_seebeck} and \ref{PbCdTe_seebeck} recover also the well known decrease of the Seebeck coefficient with the concentration of holes,\cite{Rogers, Pei2} which  is due
to  the  strong  interdependence  between  different  properties of the thermoelectric material  via  the carrier concentration.
%---------------------------------Seebeck-PbCdTe--------------------------------
%---------------------------------------------------------------------------------------
%---------------------------------Seebeck-PbMnTe--------------------------------
%---------------------------------------------------------------------------------------

\begin{acknowledgments}
The authors  thank T. Story for getting us involved in this subject and for many valuable discussions. The work was supported by the European Union within the European Regional Development Fund, through grant Innovative Economy (POIG.01.01.02-00-108/09), the U.S. Army Research Office under Contract/Grant Number W911NF-08-1-0231, and by the Polish Ministry of Science and Higher Education project (N N202 483539). The computations were carried out in the Academic Computer Center CI TASK in Gdansk.
\end{acknowledgments}

\end{document}